\documentstyle[aps,amsfonts,twocolumn,prl,floats,epsfig]{revtex}     

\begin{document}

\title{Phase transition in the collisionless damping regime for wave-particle
interaction}
\author{Marie-Christine Firpo\cite{byline} and Yves Elskens\cite{byline}}
\address{Equipe turbulence plasma de l'UMR 6633, CNRS--Universit\'e de Provence,\\
case 321, Centre de Saint-J\'er\^ome, F-13397 Marseille cedex 20}
\date{preprint TP99.10 - to appear in {\bf Physical Review Letters}}
\maketitle

\begin{abstract}
Gibbs statistical mechanics is derived for the Hamiltonian system
coupling self-consistently a wave to $N$ particles. This
identifies Landau damping with a regime where a second order phase
transition occurs. For nonequilibrium initial data with warm
particles, a critical initial wave intensity is found~: above it,
thermodynamics predicts a finite wave amplitude in the limit $N\to
\infty $~; below it, the equilibrium amplitude vanishes.
Simulations support these predictions providing new insight on the
long-time nonlinear fate of the wave due to Landau damping in
plasmas.

\end{abstract}


\unitlength 1mm

Landau damping is a striking phenomenon first evidenced in plasma
physics, but also lately in various systems such as nonlinearly
coupled oscillators \cite{Strogatz92} or neutral rarefied gases
\cite{Stubbe}. In plasma physics, its classical treatment involves
a continuous media analysis through Vlasov-Poisson equations. It
was shown by O'Neil that (near-)resonant particles play a major
role in the time-asymptotic, generically nonlinear regime
experienced by kinetic systems submitted to Landau damping
\cite{ONeil65}. This has been debated recently by several works.
In particular, Isichenko argued for algebraic final decay of the
electric field in the nonlinear regime \cite{Isi}, under the
crucial assumption of an asymptotically vanishing field
\cite{Lancomment}, though in numerical investigations by Manfredi
\cite{Manfredi} the electric field was seen to damp exponentially
in the linear regime and then to oscillate around a finite
amplitude indicating a Bernstein-Greene-Kruskal (BGK) equilibrium.
Lately, Lancellotti and Dorning \cite{LancDorn} have shown that
both types of time-asymptotic states were eligible depending on
initial conditions.

In this Letter we address this issue from a different viewpoint
and relate both possible time-asymptotic states (with zero or
finite field) to different phases in a phase transition picture.
Rather than using the large number of weakly interacting particles
$N$ to express the dynamics by a kinetic equation for a
distribution function $f(x,p)$, we note that the intrinsic chaos
in this dynamics ensures an effective ergodicity allowing a
Gibbsian equilibrium statistical mechanics treatment. This
analysis provides further insight on the results of the kinetic
approach.

As real plasmas are made of a finite number of particles and as
these graininess properties, and notably the existence of
spontaneous emission, are discarded by their classical vlasovian
treatment, a model taking these properties into account has been
derived \cite{Antoni98,Zekri96}. The first steps in this direction
were made by O'Neil \textit{et al.} \cite{ONeil71} in order to
treat the cold beam-plasma instability, and a Hamiltonian
formalism was first introduced by Mynick and Kaufman
\cite{Mynick78}. This Hamiltonian system models the interaction
between $M$ (bulk) Langmuir waves and $N$ quasiresonant (tail)
particles and has already provided insight into linear Landau
growth and damping, the cold beam-plasma instability
\cite{Tennyson94} and Van Kampen modes \cite{Zekri96}. From the
physical viewpoint, this approach has the advantage to treat the
waves as dynamical objects rather than subordinate phenomena, i.e.
excitations of the bulk plasma. In the kinetic limit $N \to
\infty$, the coupled (many-body) wave-particle dynamics approaches
the corresponding dynamics generated by the Vlasov equation over
any finite time interval \cite {kinetic}.

For simplicity, we concentrate on the single-mode case ($M=1$),
which is paradigmatic to study Landau damping. Besides, this case
is of special interest for devices such as traveling wave tubes
where a single mode can be selected \cite{TsuDo87}. For $N$
identical (tail) particles moving on the interval of length
$L=2\pi$ with periodic boundary conditions
${\mathbb{T}}:={\mathbb{R}}/(2\pi {\mathbb{Z}})$, with unit mass
and charge, and respectively position $x_{r}$ and momentum
$p_{r}$, interacting with one wave of natural frequency
$\omega_{0}$, unit wave number, phase $\theta$ and intensity $I$,
the Hamiltonian reads
\begin{equation}
H= \sum_{r=1}^{N} \frac{p_{r}^2}{2} + \omega_0 I -\sqrt{\frac{\eta \omega_0^3%
}{N}}\sum_{r=1}^{N} \sqrt{2I}\cos(x_{r}- \theta) \text{.}  \label{ham}
\end{equation}
The first term describes ballistic motion of the particles, the
second the oscillations of the free wave (harmonic oscillator) and
the third couples them; here $\eta$ is a small parameter denoting
the ratio of the tail density over the plasma density. The bulk of
the plasma is considered as a linear dielectric supporting plasma
waves. The wave frequency $\omega_{0}$ (which is the plasma
frequency $\omega_{{\rm p}}$) provides the natural reference time
scale, which can be fixed with no loss of generality. Thus
$\omega_{0}^{-1}$ will be the time unit, so that the phase
velocity of the free wave is also 1.

The self-consistent system can be viewed as a set of nonlinear oscillators
coupled through a mean-field, as the equation of motion for any particle $r$
obtained from (\ref{ham}) reads
\begin{equation}
\ddot{x}_{r} = - \sqrt{2\eta I /N} \sin(x_{r}-\theta)  \label{pendu}
\end{equation}
i.e. the equation of a pendulum in the field with strength
$\sqrt{2\eta I/N}$ and angle $\theta$. The wave evolves
consistently with the particles. For this mean-field model, the
kinetic vlasovian limit \cite{kinetic} corresponds to taking the
limit $N \to \infty$ while the acceleration felt by any particle
should remain finite. This induces the natural kinetic scaling for
the wave intensity $I={\cal O}(N)$ and gives its kinetic
equivalent as the intensive quantity $\psi:=I/N$.

Most importantly, this approach enables an equilibrium statistical
mechanics treatment, provided the dynamics is effectively ergodic.
In this Letter we derive the equilibrium Gibbs predictions for one
field's observable, its intensity $I$. This study predicts a
second-order phase transition in the regime corresponding to
damping. We exhibit a critical initial field intensity, above
which particle trapping induces a finite field amplitude, whereas
below it, in the kinetic limit of an infinite number of particles,
the asymptotic field would vanish. These results complement the
dynamical ones \cite{ONeil65,LancDorn}, and our estimations are
fully supported by
numerical simulations based on a fourth-order symplectic scheme \cite{Cary93}%
.

The Liouville sub-manifold of the phase space is fixed by the two usual
constants of the motion, total energy $E=H$ and total momentum $%
P=\sum_{r=1}^{N} p_{r}+I$. To simplify calculations, we turn to
the canonical ensemble, relying on the equivalence of ensembles
when systems get large \cite{Huang}, and introduce the temperature
$T$ ($k_{B}=1$) of the tail particles--wave system. The canonical
measure (constrained by constant momentum $P$) reads
\begin{equation}
d\mu_c = e^{-H/T} \delta \left(P-\sum_{r=1}^{N}p_{r}-I \right) dId\theta
\prod_{r=1}^{N}dp_{r}dx_{r}  \label{dmu}
\end{equation}
where the phase space variables $\left({\bf p},I,{\bf x},\theta \right) $
evolve in $\Lambda ={\mathbb{R}}^{N}\times {\mathbb{R}}^{+}\times {\mathbb{T}%
}^{N}\times {\mathbb{T}} $. With the intensive variable $\sigma :=P/N$, the
canonical partition function $Z_{c}(T,\sigma,N)$ reads
\begin{equation}
Z_{c} = (2\pi)^{\frac{3N+1}{2}} T^{\frac{N-1}{2}} N^{\frac{1}{2}}
\int_{0}^{\infty }\exp[ -\frac{N}{T}f(\sigma ,T,\psi)] d\psi
\label{defZc}
\end{equation}
where $f(\sigma ,T,\psi ):=(\sigma -\psi)^{2}/2 + \omega _0 \psi - T \ln
I_0( \sqrt{2\eta \psi }/T) $~; $I_{n}$ denotes the modified Bessel function
of order $n$. For large $N$, with $\varphi :=\sqrt{2\eta \psi }/T$, one
finds the minimum of $g(\sigma ,T,\varphi ):=f[\sigma ,T,
T\varphi^2/(2\eta)] $, by solving the saddle-point equation
\begin{equation}
\partial_{\varphi }g = \varphi T \left[ \frac{T}{\eta } \left(
\omega_{0}-\sigma +\frac{T^{2}\varphi ^{2}}{2\eta }\right) -\frac{%
I_{1}(\varphi )}{\varphi I_{0}(\varphi )}\right] =0 \text{.}  \label{defmin}
\end{equation}
The minimum of $g$ is reached at $\varphi =0$ iff $\sigma <\omega_{0}$ and $%
T \geq T_{c}$, where
\begin{equation}
T_{c} := \frac{\eta }{2\left( \omega_0-\sigma \right)} \text{.}
\label{defTc}
\end{equation}
Otherwise the minimum is reached for a certain $\varphi^{*}(\sigma
,T)
> 0$, defined implicitly by equation (\ref{defmin}). The canonical
averages for the wave intensity are respectively
\begin{eqnarray}
\left. \langle I \rangle _{c}\right| _{\varphi =0} = \frac{T^{2}}{%
(\omega_{0}-\sigma)(T-T_{c})}  \label{Iint} \\
\left. \langle I \rangle_{c} \right|_{\varphi^{*}(\sigma,T)\neq 0}= N\frac{%
T^{2}\varphi ^{*2}}{2\eta } \text{.}  \label{Iext}
\end{eqnarray}
Consequently, when $\sigma < \omega_0$, a second order phase
transition \cite{Huang,noteTrans} occurs at $T_{c}$, with
\textit{order parameter} $\psi = I/N$, i.e. the suitably
normalized intensity of the wave \cite{kinetic}. Above $T_{c}$,
the intensity of the wave loses its extensivity~: the particles
and the wave decouple in the sense that the wave controls just one
among $N+1$ degrees of freedom. On the contrary, for $T<$ $T_{c}$,
the wave is macroscopic and large enough to trap particles. When
$T \to T_{c}$, (\ref{Iint}) exhibits a formal divergence signaling
the crossover from intensive to extensive scaling (\ref{Iext}).

Finally we express the equation of state of the particles-wave system in
terms of the Gibbs average of the energy density $h:=H/N$, using $\langle H
\rangle_{c }= T^{2}\partial _{T}\ln Z_{c}$. Thus, for $\varphi = 0$,
\begin{equation}
\left. \langle h \rangle_{c}\right|_{\varphi=0} = \sigma ^{2}/2 + T/2 \text{,%
}  \label{hcphizero}
\end{equation}
where $\sigma^{2}/2$ is interpreted as the mean centre-of-mass kinetic
energy of the particles-wave system and $T/2$ as the thermal agitation
energy in this frame. For $\varphi^* \neq 0$, in the limit of large $N$, one
obtains similarly
\begin{equation}
\left. \langle h \rangle_{c} \right|_{\varphi^{*}(\sigma ,T)\neq 0} =
g\left[ \varphi ^{*}(\sigma ,T)\right] + T/2 \text{.}  \label{hcphistar}
\end{equation}

Before considering the implications of our results, let us discuss
the conditions under which they hold. The Gibbs Ansatz assumes
that particles explore the available $(x,p)$ space thanks to the
nonintegrability of their dynamics. For physical applications, it
suffices that they explore a `large' part of the $(x,p)$ space ;
in our case, that they be able to wander, say, within a velocity
range of two standard deviations on either side of their average
velocity. The constraint $\sum_{r}p_{r}+I=P$ ensures that the
kinetic energy contribution to (\ref{dmu}) is a Maxwell
distribution centered on $\sigma -\psi $, with standard deviation
$T^{1/2}$ for $p_{r}$. One therefore expects some good mixing
properties, due to a trapping/detrapping mechanism
\cite{crossings}, if $T^{1/2}$\ is of the order of the velocity
width of resonance (\ref{pendu}), namely $(\eta \psi )^{1/4}$.
This equivalently corresponds to $\varphi $\ being of order 1. In
discussing numerical simulations, we shall note that relaxation
towards equilibrium occurs for initial data roughly realizing this
condition.

Now we turn to the manifestation of this phase transition in the
wave-particle dynamics. The problem raised by Landau damping is an initial
value problem (hence a nonequilibrium one), where a packet of particles,
spatially homogeneous with velocities distributed according to a certain $%
f_{0}(v)$, interacts with a wave launched with intensity $\psi_0 > 0$. This
determines $E/N = \langle v^{2} \rangle_0/2 + \omega_0\psi_0$ and $\sigma =
\langle v \rangle_0 + \psi_0$. The condition $\sigma < \omega_{0}$ implies $%
\langle v \rangle_0 < \omega_0$, which corresponds typically to a bunch of
particles, centered on the phase velocity of the wave, initially distributed
according to a decreasing $f_0(v)$ and thus inducing Landau damping on the
wave. The initial particles temperature is $T_{0}:= \langle v^{2}
\rangle_{0} - \langle v \rangle_0^{2}$.

Given $E/N$ and $\sigma$, the equation of state (\ref{hcphizero}) in the
`high temperature' regime expresses the equilibrium temperature $T$ in terms
of initial data
\begin{equation}
T = T_0 + 2 \psi_0 ( \omega_0 - \langle v \rangle_0 ) -\psi_0^2 \text{.}
\label{Temp}
\end{equation}
Using (\ref{defTc}), defining $a:=\omega_0 - \langle v \rangle_0 > \omega_0
- \sigma > 0$ and putting $\alpha_{\pm} := a \pm \sqrt{a^{2}+T_{0}}$,
condition $T > T_c$ reduces to $2 ( \psi_0 - a) ( \psi_0 - \alpha_{+}) (
\psi_0 - \alpha_{-}) > \eta $. Therefore, provided the (near-resonant, tail)
particle distribution is warm enough ($2aT_{0}>\eta $), if $\psi_0<\psi _{0c}
$ in the limit of large $N$, the equilibrium field $\langle \psi \rangle_{c}$
vanishes, whereas for $\psi_0>\psi _{0c}$ the equilibrium field remains
finite. The critical initial wave amplitude $\psi_{0c}$ is defined by
\begin{equation}
2\left( \psi _{0c}-a\right) \left( \psi _{0c}-\alpha _{+}\right) \left( \psi
_{0c}-\alpha _{-}\right) =\eta \text{.}  \label{criteq}
\end{equation}
Moreover, whatever the initial conditions, one deduces that if $\psi_0>a$
then the asymptotic field will always be finite.

These results are illustrated for an initial normalized, warm particle distribution $%
f_{0}(v)=C_{1}-C_{2}/(2-v)$, with $C_{1},C_{2}>0$, over the range $%
v_{1}<v<v_{2}$ (with $v_{2}-\omega_0=\omega_0-v_{1}$) and $%
f_{0}(v)=0$ outside of it. The linear Landau
rate \cite{Zekri96} for amplitude $\sqrt{\psi }=\sqrt{I/N}$ is $%
\gamma_{{\rm L}}=(\pi /2)\omega_0 \eta f_0^{\prime }(\omega_0)$.
In our dynamical
simulations, $\omega_0=1$, $\eta =5.02\cdot 10^{-4}$, $v_{1}=0.75$, $%
v_{2}=1.25$ and $\gamma_{{\rm L}}=-1/200$ (from which $C_{1}$ and
$C_{2}$ are deduced). Then $\langle v \rangle_0=\allowbreak .931$
and $\langle v^{2}\rangle_0=\allowbreak .882$, thus
$T_{0}=1.50\cdot 10^{-2}$. The resulting critical mean intensity
is $\psi_{0c}=5.59\cdot 10^{-2}$.

Running simulations for different values of $\psi_0$ and $N$
yields the plot of Fig.~\ref{fig001} for asymptotic values of
$\psi $. This figure shows a clear agreement with thermodynamics
predictions in the kinetic limit $N \to \infty $ for both
quasi-ballistic and trapping regimes. In the critical region where
$\psi_0 \ {\ _{\sim}^{<}} \ \psi_{0c}$ (dashed lines), one
observes strong metastability effects and critical slowing-down,
whereas additional finite-$N$ corrections smooth the transition~:
in this region, initial field intensities are far larger than
their equilibrium averages. Moreover, for the range of $\psi_0$
considered on Fig.~\ref{fig001}, the analytic expression for $T$
happens to differ only slightly from $T_0$. Therefore the `mixing'
condition corresponds to $\sqrt{2\eta \psi_0}/T_{0}$\ being of
order 1.

This mixing condition is roughly fulfilled for the runs of
Fig.~\ref{fig002} displaying the temporal evolution of $\ln \sqrt{
\psi }$ and progress towards equipartition for different values of
$N$. Along these runs, the relative energy variation is less than
$2 \cdot 10^{-7}$. On the inset, after the initial linear Landau
damping at the prescribed rate, one observes trapping oscillations
that could suggest the setting up of a BGK equilibrium. However,
on longer times, the sweeping of the phase space is sufficient to
drive the system to a far different, `high' temperature, regime
where $I$ is no longer extensive. Critical slowing-down reflects
on the relaxation times diverging with $N$. Analogous pathological
relaxation properties in the limit $N \to \infty$ have been
reported for a similar mean-field model \cite{Latora98} in the
subcritical energy domain where chaos is paradoxically shown to be
maximal \cite{Latora98,Firpo98}.

It is surprising to note the good ergodic behavior of $\psi $ for
the initial domain where $\psi_0 \ll \psi_{0c}$, where no mixing
can be expected as the system is close to integrability. Here one
recovers the effective linear Landau damping and ergodicity is
ensured dynamically by a perturbative analysis
\cite{Zekri96,noteKac}.

In the trapping regime, the equilibrium level is quickly reached
as a substantial fraction of particles is liable to undergo
separatrix crossings \cite{crossings}. Nevertheless $\psi_0$ is
limited by the velocity range chosen for the tail particles
\cite{vrange}. One checks that the trajectories corresponding to
the velocity borders of our distribution in the Boltzmann $(x,p)$
space correspond to KAM tori for the effective 1.5 degrees of
freedom Hamiltonian dynamics (\ref{pendu}) with time dependent
parameters $(\theta ,I)$ \cite{Firpo99}.

Finally, denoting by $\tau _{{\rm B0}}=(2\eta \psi_0)^{-1/4}$ the
trapping time corresponding to the initial mean wave intensity and
putting $q=|\gamma _{{\rm L}}|\tau _{{\rm B0}}$, it turns out that
the phase transition occurs here for $q=q_{{\rm c}}\simeq 0.06$.
This agrees with the qualitative dynamical threshold drawn (in a
non self-consistent way) by O'Neil \cite{ONeil65} separating a
regime dominated by trapping (if $q<q_{{\rm c}}$) from a regime
where linear predictions are effective (if $q>q_{{\rm c}}$).
Existing numerical self-consistent simulations of the damping of a
single wave are limited to the \textit{early stage} of wave
evolution and would typically infer a larger $q_{{\rm c}}$ (e.g.
$q_{{\rm c}}=0.77$ for a linear $f_{0}(v)$ in \cite{Sugihara72}).
As clearly accounted for by Fig.~\ref{fig002} and inset (for which
$q=0.14$) , our new approach relates this apparent discrepancy to
critical slowing-down. Our intuition is that an algebraic decay of
the order parameter $\psi$ (in the limits $N \to \infty$ and $t
\to \infty$) may be a dynamical signature of the vicinity of
critical point, as in Landau second order phase transitions'
picture. In a parameter regime $q\ _{\sim}^{>} 0.06$, Isichenko's
conclusions \cite{Isi} may then apply here.

In conclusion, using the self-consistent wave-particle Hamiltonian
(\ref{ham}) to describe this system has revealed a new phenomenon,
the phase transition associated with the Landau damping regime.
This phenomenon eludes the usual Vlasov-Poisson description, which
actually treats all particles on the same footing, as a Coulomb
system. But the wave-particle interaction is effectively very
different for near-resonant particles (trapping time scale $(2\eta
\psi )^{-1/4}$ and Landau linear scale $|\gamma _{{\rm L}}|^{-1}$)
and for bulk particles (adiabatic or non-resonant averaging time
scales, exponentially longer than resonant ones): our
self-consistent Hamiltonian describes the system on the physical
time scales for the wave evolution \cite{Antoni98}.

Moreover, our finite-$N$ approach shows that, in a real plasma,
the wave will never damp completely but will eventually fluctuate
around a finite $N$-dependent thermal level due to spontaneous
wave emission.

Stimulating discussions with D.F.~Escande, F.~Doveil, P.~Bertrand
and M.~Poleni, and critical reading of the manuscript by
D.F.~Escande are gratefully acknowledged. This work is part of the
european network {\it Stability and universality in classical
mechanics} and CNRS GdR {\it Syst{\`e}mes de particules
charg{\'e}es} SParCh.




\clearpage


\begin{figure}
  \vskip 4.5cm
  \centerline{
  }
\caption{ Time averages of normalized intensity $\psi $ reached
for long times versus $\psi_0$ for $N=16000$ (triangles),
$N=32000$ (squares) and $N=64000$ (stars). Filled symbols for
simulations times of $2\cdot 10^{6}\omega_0^{-1}$. Open symbols
for run times of $5\cdot 10^{5}\omega_0^{-1}$. Canonical
estimations are drawn in continuous line for $N=16000$, 32000 and
64000 particles. Near the transition, dashes indicate that
estimate (\ref{Iint}) is not expected to be accurate.}
\label{fig001}
\end{figure}

\begin{figure}
  \vskip 4.5cm
  \centerline{
  }
\caption{ Time evolution of normalized amplitude
$\sqrt{\psi}=\sqrt{I/N}$ for $N=16000$ (dots), $N=32000$ (thick
line) and $N=64000$ (thin line), for $\psi_0 = 1.56 \cdot 10^{-3}
< \psi_{0c}$. Inset : initial evolution, including brief initial
linear damping stage. On longer times, for $N=32000$, $\psi$
wanders around half the value associated to $N=16000$, in
agreement with canonical predictions. Relaxation towards
equilibrium is much longer for $N=64000$ due to critical
slowing-down near the phase transition.} \label{fig002}
\end{figure}

\clearpage

\begin{figure}
  \centerline{
  \psfig{figure=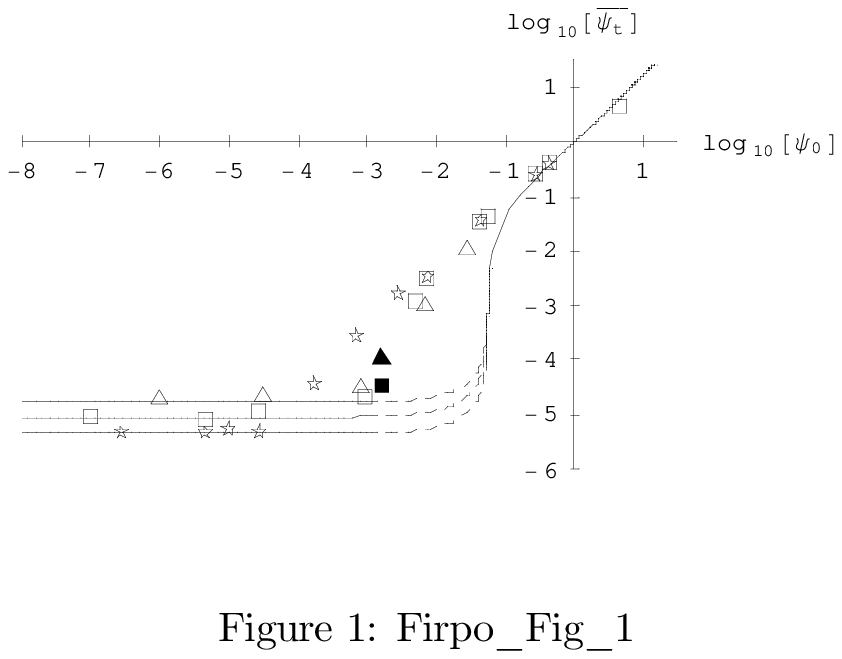,width=8cm,height=4cm}
  }
\end{figure}

\begin{figure}
  \centerline{
  \psfig{figure=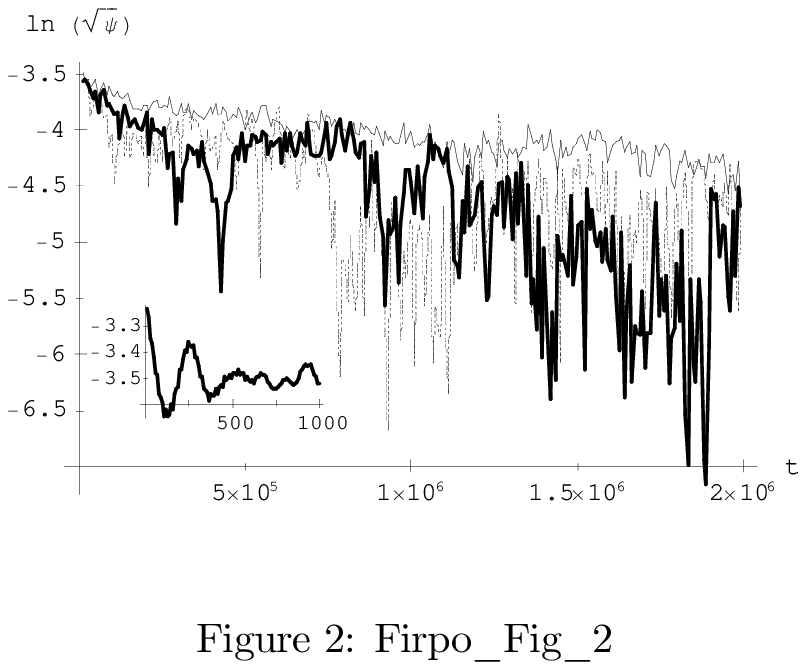,width=8cm,height=4cm}
  }
\end{figure}


\begin{references}
\bibitem{noteTrans} {It is important to note that the
long range nature of the wave-particle interaction makes it
possible to observe a phase transition even in a one
space-dimension system.}
\bibitem{noteKac}  {Ergodicity for the collective variable $\psi $ in this
domain can also been accounted for using a result of Kac on the distribution
of sums of trigonometric functions \cite{Firpo99}.}
\bibitem{vrange}  {An additional motivation for limiting the velocity range
of the tail particles distribution is that one does not want the
resonant region to extend as far as the bulk of the plasma, as
this would also break the validity of the self-consistent
Hamiltonian model \cite{Antoni98}.}
\end{references}
\end{document}